\documentclass[aps,pra,twocolumn,showpacs,floatfix]{revtex4}
\usepackage{graphicx}% Include figure files
\usepackage{dcolumn}% Align table columns on decimal point
\begin{document}
\title{Third-order relativistic many-body calculations of energies
and lifetimes of levels along the silver isoelectronic sequence}
\author{U. I. Safronova}
\affiliation{Department of Physics, 225 Nieuwland Science Hall\\
University of Notre Dame, Notre Dame, Indiana  46566}
\author{I. M. Savukov}
\affiliation{Department of Physics, Princeton University,
 Princeton, New Jersey 08544}
\author{M. S. Safronova}
 \affiliation {Electron and Optical Physics
Division,\\ National Institute of Standards and Technology,
Gaithersburg, Maryland, 20899-8410}
\author{W. R. Johnson}
\affiliation{
 Department of Physics, 225   Nieuwland Science Hall\\
 University of Notre Dame, Notre Dame, Indiana 46566}
\date{\today}
\begin{abstract}
Energies of $5l_j$ ($l= s,\, p,\,d,\, f,\, g$) and $4f_j$ states in neutral Ag and Ag-like ions with
nuclear charges $Z = 48 - 100$ are calculated using relativistic many-body
perturbation theory.
Reduced matrix elements, oscillator strengths, transition rates
and lifetimes are calculated for the 17 possible $5l_j-5l^{\prime}_{j'}$ and
$4f_j-5l_{j'}$ electric-dipole transitions.
Third-order corrections to energies and dipole matrix elements are
included for neutral Ag and for ions with $Z \leq 60$. Second-order
corrections are included for $Z>60$.
Comparisons are made with available experimental data for transition energies
and lifetimes.
Correlation energies and transition rates are shown graphically as functions
of nuclear charge $Z$ for selected cases.
These calculations provide a theoretical benchmark for comparison with experiment and theory.
 \pacs{31.15.Ar, 32.70.Cs, 31.25.Jf, 31.15.Md}
\end{abstract}
\maketitle
\section{Introduction}
This work continues earlier third-order relativistic many-body
perturbation theory (RMBPT) studies of energy levels of ions with
one valence electron outside a closed core. In Refs.~\cite{li-en,na-en,cu-en} 
third-order RMBPT was used to calculate
energies of the three lowest states ($ns_{1/2}$, $np_{1/2}$, and $np_{3/2}$)
in Li-, Na-, and Cu-like ions along the respective isoelectronic
sequences, while in the present work, third-order RMBPT is used
to calculate energies of the eleven lowest levels, $5s_{1/2}$,
$5p_j$, $5d_j$, $4f_j$, $5f_j$, and $5g_j$ in
Ag-like ions. It should be noted that the $n=1,\, 2,\, \text{and } 3$ cores of
Li-, Na-, and Cu-like ions are completely filled, by contrast with Ag-like ions,
where the $n=4$ core [Cu$^+$]$4s^24p^64d^{10}$ is incomplete.

Third-order RMBPT calculations of $5s_{1/2}-5p_j$ transition amplitudes
in Ag-like ions up to $Z$=60 were previously performed by
\citet{chou}. In the present paper, we extend the calculations of \cite{chou}
to obtain energies, reduced matrix elements,
oscillator strengths, and transition rates for the 17 possible
$5l_j-5l^{\prime}_{j'}$ and $4f_j-5l_{j'}$ E1 transitions.
Additionally, we evaluate lifetimes of excited states.
Most earlier theoretical studies of Ag-like ions were devoted
to oscillator strengths and lifetimes
\cite{martin,migd} rather than energy levels; an exception is
the work of \citet{kim} in which energies, oscillator strengths
and lifetimes of levels in Ag-like ions were calculated
using relativistic Dirac-Fock (DF) wave functions  \cite{des}.
In the present paper, we use RMBPT to determine energies and lifetimes
of $4f_j$ and $5l_j$ levels in neutral Ag and Ag-like ions with $Z = 48 - 100$.
We  compare our results with experimental data from
Refs.~\cite{ag1,cd1,cd2,49,50,51,52,53,sugar}.

\begin{table*}
\caption{\label{tab1} Contributions to energy levels of
Ag-like ions  in cm$^{-1}$.}
\begin{ruledtabular}
\begin{tabular}{lrrrrrrrcrrrrrrr}
\multicolumn{1}{c}{$nlj$ } &
\multicolumn{1}{c}{$E^{(0)}$} &
\multicolumn{1}{c}{$E^{(2)}$} & \multicolumn{1}{c}{$E^{(3)}$} &
\multicolumn{1}{c}{$B^{(1)}$} &
\multicolumn{1}{c}{$B^{(2)}$} & \multicolumn{1}{c}{$E_{\rm LS}$}&
\multicolumn{1}{c}{$E_{\rm tot}$} &
\multicolumn{1}{c}{} &
\multicolumn{1}{c}{$E^{(0)}$} &
\multicolumn{1}{c}{$E^{(2)}$} & \multicolumn{1}{c}{$E^{(3)}$} &
\multicolumn{1}{c}{$B^{(1)}$} &
\multicolumn{1}{c}{$B^{(2)}$} & \multicolumn{1}{c}{$E_{\rm LS}$}&
\multicolumn{1}{c}{$E_{\rm tot}$} \\
\hline
\multicolumn{8}{c}{$Z$=47 } &
\multicolumn{1}{c}{       } &
\multicolumn{7}{c}{$Z$=48 }\\
$5s_{1/2}$&  -50376& -11173& 3139&   83& -180& 138& -58369&& -124568& -13505& 3321&  150& -264& 162&-134703\\
$5p_{1/2}$&  -26730&  -4217&  680&   34&  -60&  -2& -30296&&  -84903&  -7731& 1614&  107& -143&  -3& -91058\\
$5p_{3/2}$&  -26148&  -3872&  617&   23&  -50&   7& -29423&&  -82871&  -7176& 1477&   76& -130&   8& -88615\\
$5d_{3/2}$&  -11982&   -328&   36&    1&  -11&  -2& -12287&&  -45147&  -1531&  236&   14&  -36&  -2& -46466\\
$5d_{5/2}$&  -11967&   -323&   34&    1&  -12&   2& -12265&&  -45010&  -1503&  230&   11&  -38&   2& -46308\\
$4f_{5/2}$&   -6860&    -37&    5&    0&    0&  -2&  -6894&&  -27540&   -361&   53&    0&    0&  -2& -27850\\
$4f_{7/2}$&   -6860&    -37&    5&    0&    0&   2&  -6890&&  -27543&   -361&   52&    0&    0&   2& -27849\\
$5f_{5/2}$&   -4391&    -33&   10&    0&    0&  -1&  -4414&&  -17644&   -221&   30&    0&    0&  -1& -17836\\
$5f_{7/2}$&   -4391&    -33&   10&    0&    0&   1&  -4412&&  -17646&   -221&   30&    0&    0&   1& -17837\\
$5g_{7/2}$&   -4389&     -6&    3&    0&    0&  -1&  -4394&&  -17559&    -49&    7&    0&    0&  -1& -17601\\
$5g_{9/2}$&   -4389&     -6&    3&    0&    0&   1&  -4392&&  -17559&    -49&    7&    0&    0&   1& -17599\\
\multicolumn{8}{c}{$Z$=53 } &
\multicolumn{1}{c}{       } &
\multicolumn{7}{c}{$Z$=54 }\\
$5s_{1/2}$& -692263& -18545& 4916&  586& -564& 327& -705543&& -839765& -19240& 5054&  698& -628& 371& -853511\\
$5p_{1/2}$& -589599& -15210& 3861&  673& -442&  -4& -600720&& -725343& -16192& 4100&  825& -508&  -4& -737122\\
$5p_{3/2}$& -575199& -14239& 3611&  483& -417&  20& -585742&& -707377& -15159& 3839&  592& -480&  23& -718562\\
$4f_{5/2}$& -419424& -16897& 3699&  197& -790&  -5& -433221&& -572051& -20265& 4554&  306&-1108&  -6& -588570\\
$4f_{7/2}$& -419399& -16631& 3629&  131& -773&   5& -433037&& -571685& -19938& 4469&  205&-1085&   5& -588028\\
$5d_{3/2}$& -425432&  -8511& 1929&  229& -272&   0& -432058&& -536495&  -9746& 2228&  298& -334&   0& -544048\\
$5d_{5/2}$& -423242&  -8370& 1895&  178& -277&   5& -429812&& -533632&  -9578& 2186&  232& -340&   6& -541127\\
$5f_{5/2}$& -264311&  -6586& 1591&   82& -305&  -2& -269531&& -350670&  -7548& 1821&  105& -345&  -2& -356640\\
$5f_{7/2}$& -264076&  -6513& 1529&   55& -301&   2& -269303&& -350271&  -7488& 1804&   72& -342&   2& -356224\\
$5g_{7/2}$& -215806&  -1723&  319&    1&   -4&  -2& -217214&& -282283&  -2442&  457&    2&   -7&  -2& -284276\\
$5g_{9/2}$& -215811&  -1722&  319&    0&   -4&   2& -217215&& -282292&  -2440&  456&    0&   -7&   2& -284280\\
\multicolumn{8}{c}{$Z$=57 } &
\multicolumn{7}{c}{$Z$=58 }\\
$5s_{1/2}$&-1345535& -21098& 5350&1092& -832& 526&-1360497 &&-1534879& -21658& 5423& 1243& -903& 586&-1550187\\
$4f_{5/2}$&-1149152& -26307& 5919& 695&-1956&  -8&-1170808 &&-1376678& -27562& 6137&  844&-2217&  -9&-1399486\\
$4f_{7/2}$&-1147163& -25901& 5820& 474&-1921&   8&-1168685 &&-1373977& -27142& 6036&  577&-2180&   9&-1396676\\
$5p_{1/2}$&-1196202& -18769& 4647&1367& -717&  -4&-1209677 &&-1373944& -19536& 4790& 1578& -791&  -4&-1387907\\
$5p_{3/2}$&-1164895& -17563& 4363& 980& -678&  36&-1177757 &&-1337210& -18278& 4500& 1129& -747&  41&-1350564\\
$5d_{3/2}$& -932064& -13156& 3000& 558& -536&   0& -942198 &&-1084072& -14211& 3221&  663& -608&   0&-1095007\\
$5d_{5/2}$& -926588& -12895& 2929& 433& -543&   9& -936655 &&-1077506& -13912& 3139&  514& -617&  10&-1088372\\
$5f_{5/2}$& -669070& -11088& 2495& 192& -475&  -4& -677951 && -794882& -12370& 2713&  230& -529&  -5& -804843\\
$5f_{7/2}$& -668137& -11029& 2479& 134& -475&   4& -677023 && -793733& -12300& 2694&  162& -529&   5& -803702\\
$5g_{7/2}$& -536613&  -5673& 1042&   0&  -27&  -4& -541275 && -639897&  -7286& 1284&    9&  -39&  -5& -645935\\
$5g_{9/2}$& -536641&  -5647& 1039&   0&  -27&   4& -541271 && -639931&  -7232& 1282&    4&  -38&   5& -645910
\end{tabular}
\end{ruledtabular}
\end{table*}
\begin{table}
\caption{\label{tab2} Contributions to the correlation  energies of $4f$ and
 $5l$ states of
Pm$^{14+}$  in cm$^{-1}$.}
\begin{ruledtabular}
\begin{tabular}{lrrrrr}
 \rule[-2ex]{0em}{5ex} & $E^{(2)}_{l\leq 6}$& $E^{(2)}$&
$E^{(2)}_{l>6}$& $E^{(3)}_{l\leq 6}$ &
$E^{(3)}_{l>6}$  \\
 \hline
$5s_{1/2}$&  -22892& -23111&   -218&  5593&  53\\
$4f_{5/2}$&  -28621& -30123&  -1501&  6454& 338\\
$4f_{7/2}$&  -28170& -29661&  -1491&  6352& 336\\
$5p_{1/2}$&  -21377& -21632&   -255&  5139&  61\\
$5p_{3/2}$&  -19996& -20232&   -235&  4837&  57\\
$5d_{3/2}$&  -16956& -17178&   -222&  3800&  50\\
$5d_{5/2}$&  -16534& -16752&   -218&  3677&  48\\
$5f_{5/2}$&  -15891& -16179&   -288&  3296&  60\\
$5f_{7/2}$&  -15763& -16050&   -287&  3265&  59\\
$5g_{7/2}$&  -17572& -17880&   -308&   434&   8\\
$5g_{9/2}$&  -16846& -17152&   -306&   788&  14
 \end{tabular}
\end{ruledtabular}
\end{table}
\begin{figure*}[tbp]
\centerline{\includegraphics[scale=0.35]{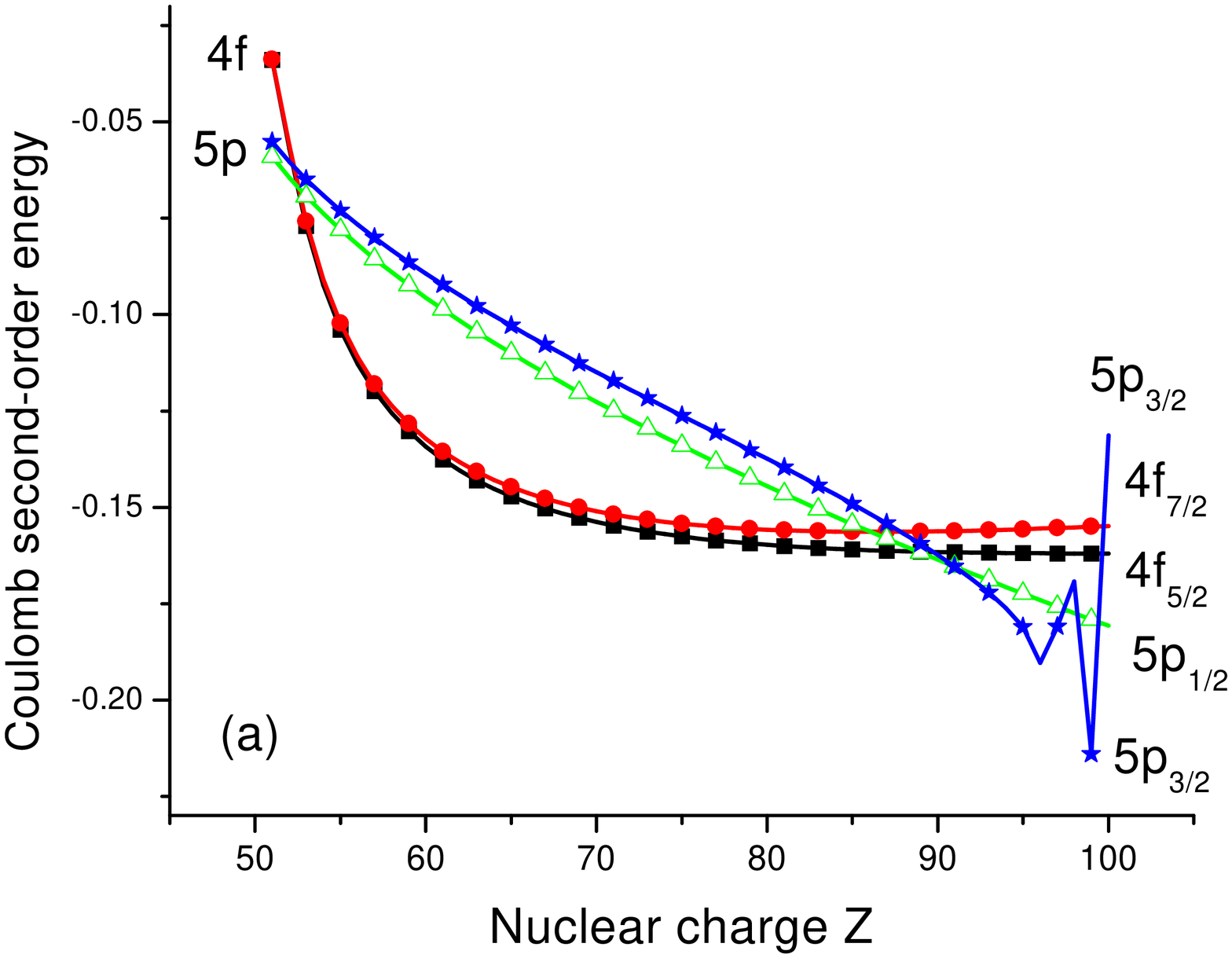}
            \includegraphics[scale=0.35]{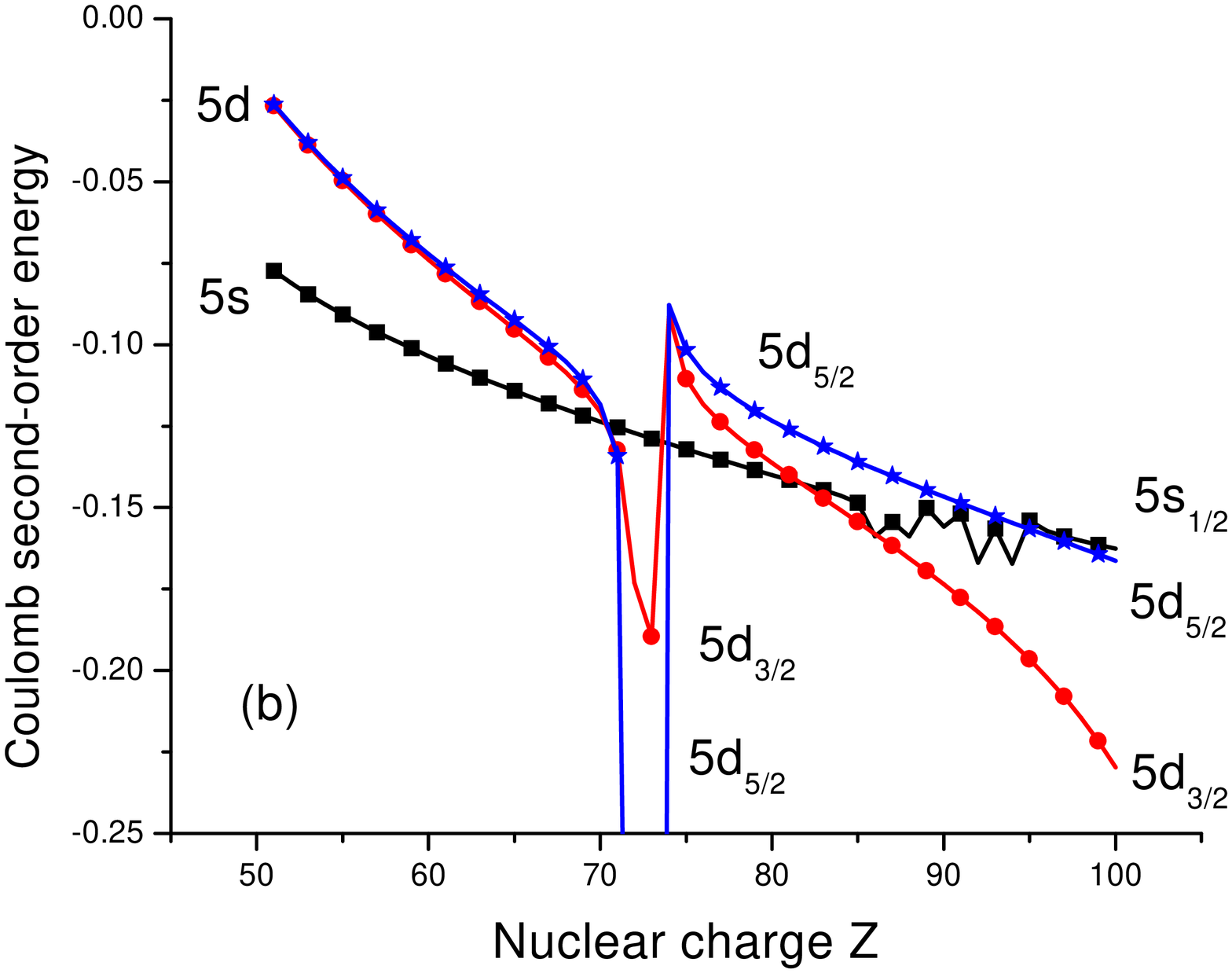}}
\centerline{\includegraphics[scale=0.35]{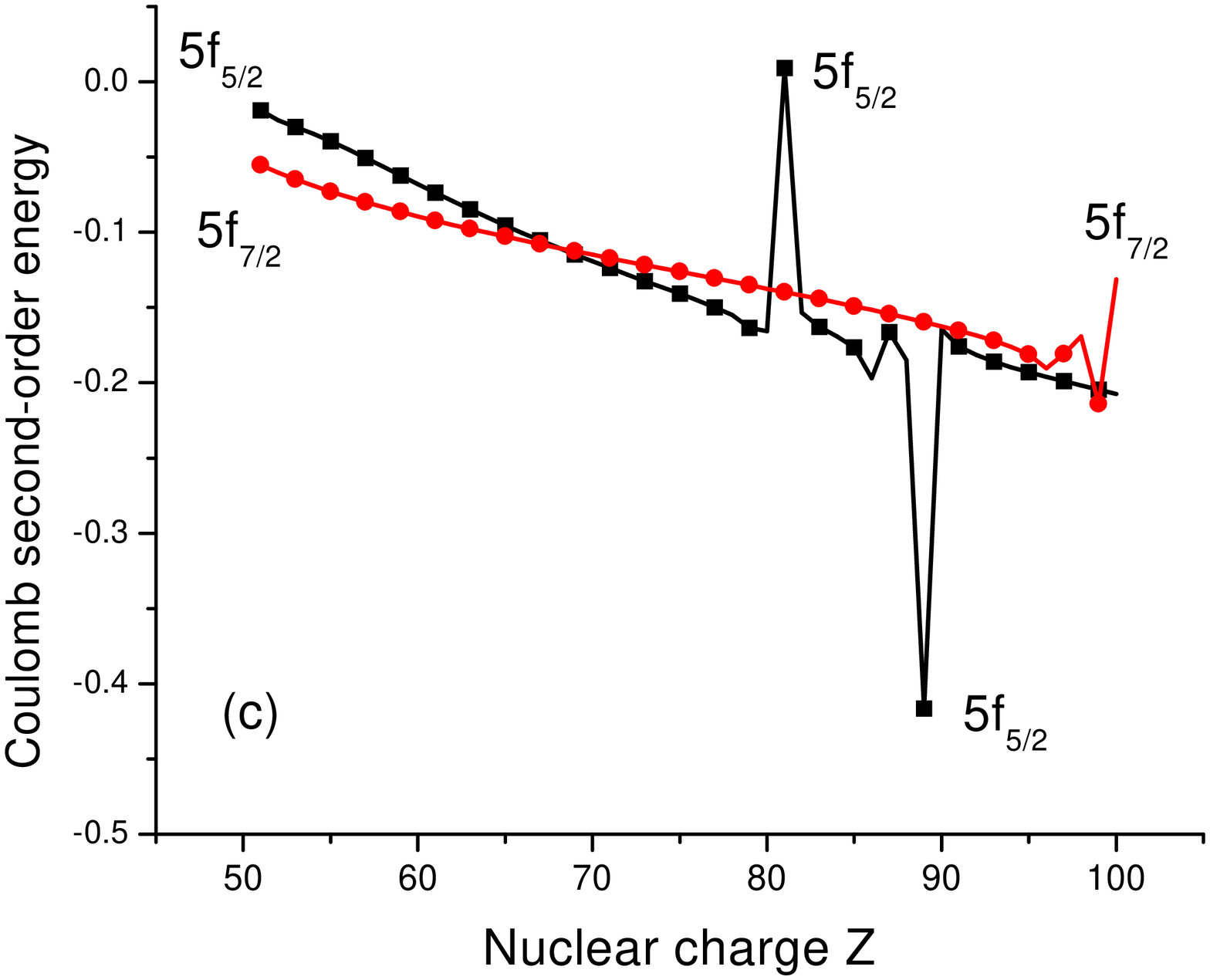}
            \includegraphics[scale=0.35]{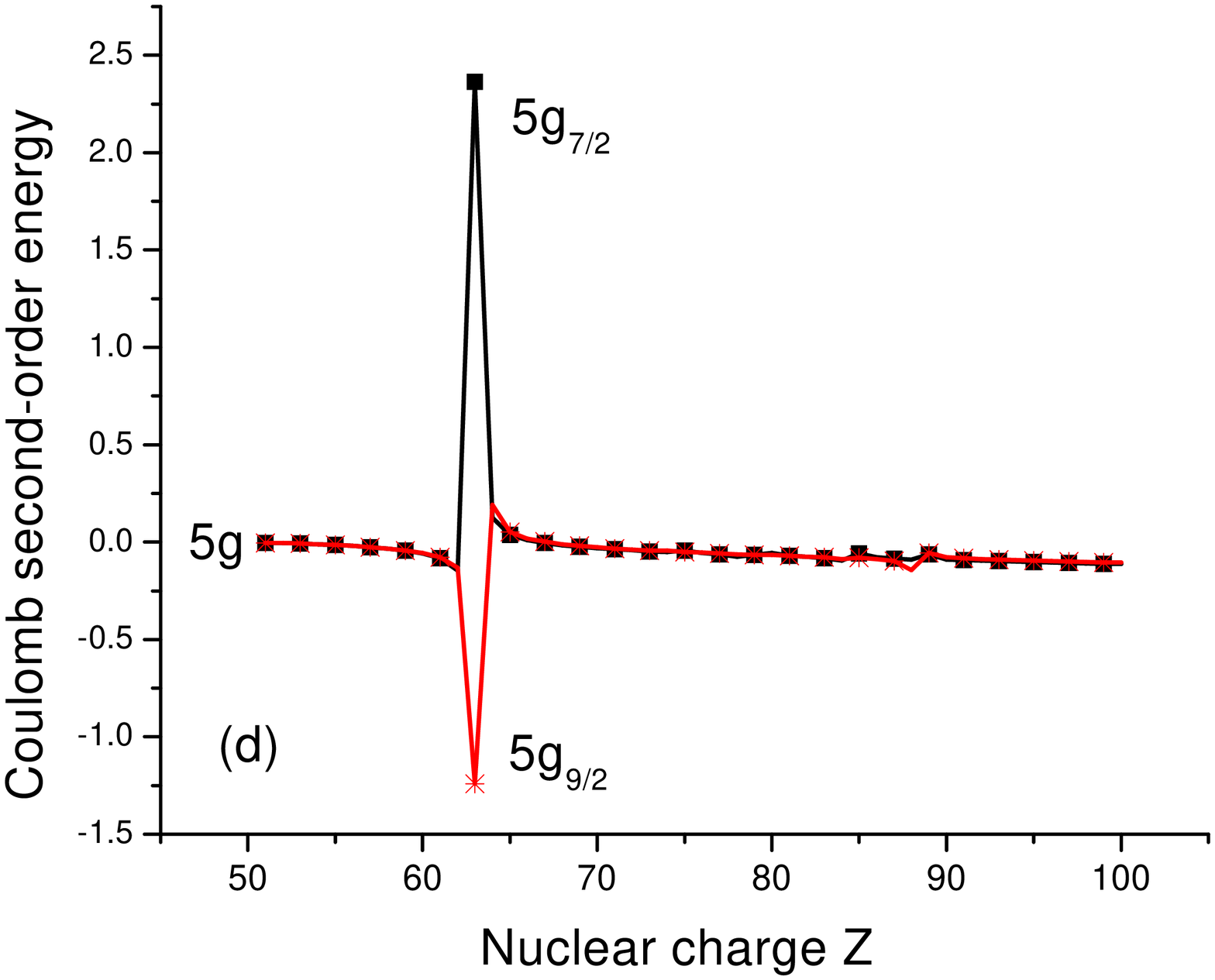}}
\caption{ Coulomb second-order contributions to valence energies
in Ag-like ions in atomic units.} \label{fig1}
\end{figure*}

\section{Energy levels of Ag-like ions}

Results of our third-order calculations of energies, which were carried 
out following the pattern
described in \cite{li-en}, are summarized in Table~\ref{tab1}, 
where we list lowest-order, Dirac-Fock
energies $E^{(0)}$, first-order Breit energies  $B^{(1)}$,
second-order Coulomb $E^{(2)}$ and Breit $B^{(2)}$ energies,
third-order Coulomb energies  $E^{(3)}$, single-particle
Lamb shift corrections $E_{\rm LS}$, and the sum of the above
$E_{\text{tot}}$. The first-order Breit energies $B^{(1)}$ include corrections
for ``frequency-dependence'', whereas second-order Breit energies
are evaluated using the static Breit operator.
The Lamb shift $E_{\text{LS}}$ is approximated as the sum of the one-electron self
energy and the first-order vacuum-polarization energy.
The vacuum-polarization contribution is calculated from the Uehling
potential using the results of Fullerton and Rinker \cite{vacuum}.
The self-energy contribution is estimated for $s$, $p_{1/2}$ and
$p_{3/2}$ orbitals by interpolating
among the values obtained by  \citet{mohr1,mohr2,mohr3} using Coulomb
wave functions. For this purpose, an effective nuclear charge $Z_{\rm eff}$ 
is obtained by finding the value of $Z_{\rm eff}$ required to give a Coulomb
orbital with the same average $\langle r\rangle$
as the DHF orbital.

We find that correlation corrections to energies in
neutral Ag and in low-$Z$ Ag-like ions are large, especially for
$5s$ states. For example, $E^{(2)}$ is 22\% of $E^{(0)}$ and
$E^{(3)}$ is 28\% of $E^{(2)}$ for the $5s$ state of neutral Ag.
These ratios decrease for the other (less penetrating) states and
for more highly charged ions.
Despite the slow convergence of the perturbation expansion, the $5s$
energy from the present third-order RMBPT calculation is within 5\%
of the measured ionization energy for the $5s$ state of
neutral Ag and improves for higher valence states and for more highly charged ions.
We include results for neutral Ag and five Ag-like ions
with $Z$ = 48, 53, 54, 57, and 58 in  Table~\ref{tab1}.
For each ion, the states are ordered by energy and
the order changes with $Z$. For example, the $4f_{5/2}$ and
$4f_{7/2}$ states are in the sixth and seventh places for neutral Ag
and Ag-like ions with the nuclear charge  $Z < 53$,
in the fourth  and fifth places for ions with $Z = 53 - 56$,
the second and third places for ions with $Z = 57 - 60$, and
the first and second places for ions with $Z \geq 62$. It
should be mentioned that the difference in energies of $5s$ and
$4f_{5/2}$ states is about 244 cm$^{-1}$ for  $Z$ = 61 which
may exceed the accuracy of the present calculations.

Below, we describe a few  numerical details of the
calculation for a specific case, Pm$^{14+}$ ($Z=61$).
We use B-spline methods \cite{Bspline} to
generate a complete set of basis DF wave functions for
use in the evaluation of RMBPT expressions.
For Pm$^{14+}$, we use 40 splines of order $k=7$ for each angular momentum. 
The basis orbitals are constrained to a cavity of
radius 10~$a_0$ for Pm$^{14+}$, where $a_0\approx0.052918$~nm is Bohr radius.
 The cavity radius is scaled for different ions;
it is large enough to accommodate all $5l_j$ and $4f_j$ orbitals considered 
here
and small enough that 40 splines can approximate inner-shell DF wave functions
with good precision.
We use 35 out 40 basis orbitals for each partial wave in our third-order
energy calculations,
since contribution of the five highest-energy orbitals is negligible.
The second-order calculation includes partial waves up to
$l_{\text{max}}=8$ and is extrapolated to account for contributions from
higher partial waves.  A lower number of partial waves,
$l_{\text{max}}=6$, is used in the third-order calculation.
We list the second-order energy
with $l_{\text{max}}=6$, the final value of $E^{(2)}$ extrapolated from $l_{\text{max}} =8$
and the contribution from partial waves with $l>6$ in
Table~\ref{tab2}. Since the asymptotic $l$-dependence of the second- and third-order
energies are similar (both fall off as $l^{-4}$), we may use the second-order
remainder as a guide to extrapolating the third-order energy, which is listed, together
with an estimated remainder from $l>6$ in the fifth and sixth columns
of Table~\ref{tab2}.
We find that the contribution to the third-order energy from states with $l>6$
is more than 300 cm$^{-1}$ for $4f$ states of Pm$^{14+}$.

\begin{table*}
\caption{\label{tab3} Comparison of the energies of the $4f$ and
$5l$ states in Ag-like ions  with experimental data  \protect\cite{nist}.
 Units: cm$^{-1}$.}
\begin{ruledtabular}
\begin{tabular}{lrrrrrrrrrrrrrr}
\multicolumn{1}{c}{$nlj$ } &
\multicolumn{1}{c}{$E_{\text{tot}}$} &
\multicolumn{1}{c}{$E_{\text{expt}}$} &
\multicolumn{1}{c}{$\delta E$} &
\multicolumn{1}{c}{$E_{\text{tot}}$} &
\multicolumn{1}{c}{$E_{\text{expt}}$} &
\multicolumn{1}{c}{$\delta E$} &
\multicolumn{1}{c}{$E_{\text{tot}}$} &
\multicolumn{1}{c}{$E_{\text{expt}}$} &
\multicolumn{1}{c}{$\delta E$} &
\multicolumn{1}{c}{$E_{\text{tot}}$} &
\multicolumn{1}{c}{$E_{\text{expt}}$} &
\multicolumn{1}{c}{$\delta E$}\\
\hline
\multicolumn{1}{c}{} &
\multicolumn{3}{c}{$Z$=47} &
\multicolumn{3}{c}{$Z$=48} &
\multicolumn{3}{c}{$Z$=49} &
\multicolumn{3}{c}{$Z$=50} \\
$5s_{1/2}$&-58369&-61106&2737&-134703&-136375&1672&-224666&-226100&1434&-327453&-328550&1097\\
$5p_{1/2}$&-30296&-31554&1258& -91058& -92239&1181&-167850&-168919&1069&-258188&-258986& 798\\
$5p_{3/2}$&-29423&-30634&1211& -88615& -89756&1141&-163546&-164577&1031&-251717&-252478& 761\\
$5d_{3/2}$&-12287&-12362&  75& -46466& -46686& 220& -97584& -97647&  63&-162915&-163245& 330\\
$5d_{5/2}$&-12265&-12342&  77& -46308& -46532& 224& -97176& -97358& 182&-162170&-163139& 969\\
$4f_{5/2}$& -6894& -6901&   7& -27850& -27956& 106& -63903& -64131& 228&-117959&-118232& 273\\
$4f_{7/2}$& -6890& -6901&  11& -27849& -27943&  94& -63922& -64123& 201&-118035&-118292& 257\\
$5f_{5/2}$& -4414& -4415&   1& -17836& -17931&  95& -41032&       &    & -76428&       &    \\
$5f_{7/2}$& -4412& -4415&   3& -17837& -17829&  -8& -41049&       &    & -76472&       &    \\
$5g_{7/2}$& -4394& -4395&   1& -17601& -17605&   4& -39656& -39578& -78& -70585& -70267&-318\\
$5g_{9/2}$& -4392& -4395&   3& -17599& -17605&   6& -39654& -39578& -76& -70581& -70268&-313
\end{tabular}
\end{ruledtabular}
\end{table*}

In Fig.~\ref{fig1}, we illustrate the $Z$ dependence of the
second-order energy $E^{(2)}$ given in atomic units for $4f$,
 $5s$, $5p$, $5d$, $5f$, and $5g$ states of Ag-like ions.
 The atomic unit (a.u.) of energy is 
  $E_h\approx4.3597\times10^{-18}$~J, where $E_h$ is Hartree
 energy. 
 
As we see from this figure $|E^{(2)}|$
slowly increases with $Z$ for most values of $Z$.
We observe several  sharp features in the curves describing $5d_{3/2}$
states ($Z=72$), $5f_{5/2}$ states ($Z=89$),
and $5g_{7/2}$ states ($Z=63$). These irregularities have their origin in the
near degeneracy of one-electron valence states with
two-particle one-hole states of the same angular symmetry, resulting in
exceptionally small energy denominators in double-excitation contributions
to the second-order energy. To remove these irregularities,  
the perturbative treatment should be based on a lowest-order wave function
that includes both one-particle and two-particle one-hole states.
The singularities in the second-order $5d_{3/2}$ energy at $Z=72$ in Fig.~\ref{fig1},
for example, occurs because the lowest-order $5d_{3/2}$ energy,
$\epsilon_{5d_{3/2}} = - 19.2215$~a.u.\
is nearly degenerate with the lowest-order energy of the $(4d_{5/2})^{-1}(4f_{5/2})^2$
state: $-\epsilon_{4d_{5/2}} +2\epsilon_{4f_{5/2}} = 36.8850-2\times 28.0685 = -19.2520$ a.u..
The other singularities seen in Fig.~\ref{fig1} have similar explanations.

\subsection*{Results and comparisons for energies}
As discussed above, starting from $Z=62$ the second-order energy
for one-particle states has
irregularities associated with nearly degenerate two-particle one-hole states.
These near degeneracies, of course, lead to similar irregularities in the
third-order valence energy.
The importance of  third-order corrections
decreases substantially with $Z$, it contributes 6\%  to the total energy
of the $5s$ state for neutral Ag but only 0.3\% to the total energy of the $5s$ state of
Ag-like Ce, $Z=58$.
Thus, omission of the third-order corrections is justified for ions with $Z>60$.
\begin{table*}
\caption{\label{tab-osc} Dipole oscillator strengths ($f$) for
transitions in Ag-like ions. }
\begin{ruledtabular}
\begin{tabular}{lllclllclllclll}
\multicolumn{1}{c}{} &
\multicolumn{2}{c}{$5s$-$5p_{j'}$ } &
\multicolumn{1}{c}{} &
\multicolumn{3}{c}{$5p_{j}$-$5d_{j'}$ } &
\multicolumn{1}{c}{} &
\multicolumn{3}{c}{$4f_{j}$-$5d_{j'}$ } &
\multicolumn{1}{c}{} &
\multicolumn{3}{c}{$4f_{j}$-$5g_{j'}$ }\\
\multicolumn{1}{c}{$Z$ } &
\multicolumn{1}{c}{1/2-1/2} &
\multicolumn{1}{c}{1/2-3/2} &
\multicolumn{1}{c}{} &
\multicolumn{1}{c}{1/2-3/2} &
\multicolumn{1}{c}{3/2-3/2} &
\multicolumn{1}{c}{3/2-5/2} &
\multicolumn{1}{c}{} &
\multicolumn{1}{c}{5/2-3/2} &
\multicolumn{1}{c}{5/2-5/2} &
\multicolumn{1}{c}{7/2-5/2} &
\multicolumn{1}{c}{} &
\multicolumn{1}{c}{5/2-7/2} &
\multicolumn{1}{c}{7/2-7/2} &
\multicolumn{1}{c}{7/2-9/2} \\
\hline
47& 0.2497&  0.5134&& 0.5773&  0.0613&  0.5491&& 1.0118& 0.0484& 0.9678&&1.3800& 0.0383&  1.3405\\
48& 0.2548&  0.5423&& 0.8097&  0.0841&  0.7540&& 1.1007& 0.0527& 1.0536&&1.2814& 0.0356&  1.2452\\
49& 0.2519&  0.5478&& 0.9113&  0.0932&  0.8387&& 1.0915& 0.0520& 1.0394&&1.2084& 0.0335&  1.1736\\
50& 0.2489&  0.5508&& 0.9577&  0.0968&  0.8736&& 0.8751& 0.0413& 0.8251&&1.0093& 0.0280&  0.9797\\
51& 0.2456&  0.5521&& 0.9781&  0.0978&  0.8856&& 0.4851& 0.0226& 0.4503&&0.7564& 0.0210&  0.7357\\
52& 0.2419&  0.5522&& 0.9846&  0.0975&  0.8854&& 0.1580& 0.0071& 0.1419&&0.5454& 0.0152&  0.5331\\
53& 0.2380&  0.5515&& 0.9833&  0.0965&  0.8785&& 0.0073& 0.0009& 0.0126&&0.4256& 0.0119&  0.4180\\
54& 0.2340&  0.5503&& 0.9773&  0.0950&  0.8677&& 0.0582& 0.0044& 0.0654&&0.3646& 0.0102&  0.3595\\
55& 0.2299&  0.5490&& 0.9684&  0.0933&  0.8546&& 0.0804& 0.0058& 0.0882&&0.3325& 0.0093&  0.3287\\
56& 0.2259&  0.5475&& 0.9580&  0.0915&  0.8403&& 0.0899& 0.0065& 0.0978&&0.3123& 0.0088&  0.3097\\
57& 0.2219&  0.5461&& 0.9467&  0.0896&  0.8254&& 0.0936& 0.0067& 0.1013&&0.2939& 0.0083&  0.2927\\
58& 0.2179&  0.5448&& 0.9350&  0.0877&  0.8103&& 0.0944& 0.0067& 0.1019&&0.2681& 0.0075&  0.2693\\
59& 0.2141&  0.5435&& 0.9232&  0.0858&  0.7955&& 0.0939& 0.0066& 0.1011&&0.2221& 0.0061&  0.2287\\
60& 0.2103&  0.5428&& 0.9118&  0.0840&  0.7805&& 0.0928& 0.0065& 0.0997&&0.1257& 0.0025&  0.1514
\end{tabular}
\end{ruledtabular}
\end{table*}

We compare our results for energy levels of
the $5l_j$ and $4f_j$ states with recommended data from the
National Institute of Standards and Technology (NIST) database
\cite{nist} in neutral Ag and Ag-like ions with $Z = 48 - 50$ in
Table~\ref{tab3}.
Although our results are generally in good agreement with the 
NIST data, discrepancies were found.
One cause for the discrepancies is that fourth- and higher-order
correlation corrections are omitted in the theory.
Another possible cause may be the large uncertainties
in the experimental ionization potentials of ions with  $Z = 49$ and 50 
in Ref.~\cite{nist}.
We also find unusually large discrepancies in the values of
$5d_{3/2} - 5d_{5/2}$ splittings for ions with $Z = 49$ and 50.
Additional tables are included in the accompanying EPAPS document
Ref.~\cite{EPAPS}, where we give energies of $5l_j$ and $4f_j$ 
states in Ag-like ions for the entire isoelectronic sequence up 
to $Z=100$.

\section{Line strengths, oscillator strengths, transition rates,
and lifetimes in Ag-like ions}
We calculate reduced electric-dipole matrix elements using the
gauge-independent third-order perturbation theory developed
in \citet{igor}. The
precision of this method has been demonstrated for alkali-metal atoms.
Gauge-dependent ``bare'' dipole matrix elements are
replaced with gauge-independent random-phase approximation (RPA) 
matrix elements to obtain
gauge-independent third-order matrix elements.
As in the case of the third-order energy,
a limited number of partial waves with $l_{\text{max} }<7$ is
included. This restriction is not very important for the ions considered here
because the third-order correction is quite small, but the truncation gives
rise to some loss of gauge invariance.
\begin{table}
\caption{\label{tab-s} Line strengths (a.u.) for E1 transitions
in Ag-like xenon ($Z$ = 54) calculated
in lowest order (DF approximation) $S^{(1)}$, second order $S^{(2)}$,
and third order $S^{(3)}$. The first-order result, which is gauge-dependent,
is given in length form. The second-, and third-order results are gauge
independent.}
\begin{ruledtabular}
\begin{tabular}{lrrr}
\multicolumn{1}{c}{Transition}&
\multicolumn{1}{c}{$S^{(1)}$ }&
\multicolumn{1}{c}{$S^{(2)}$ }&
\multicolumn{1}{c}{$S^{(3)}$ }\\
\hline
$5s_{1/2}-5p_{1/2}$&  1.876&  1.263&  1.324\\
$5s_{1/2}-5p_{3/2}$&  3.774&  2.564&  2.685\\
$5p_{1/2}-5d_{3/2}$&  4.399&  3.279&  3.333\\
$5p_{3/2}-5d_{3/2}$&  0.939&  0.707&  0.717\\
$5p_{3/2}-5d_{5/2}$&  8.416&  6.350&  6.439\\
$4f_{5/2}-5d_{3/2}$&  3.251&  2.734&  2.583\\
$4f_{5/2}-5d_{5/2}$&  0.229&  0.194&  0.182\\
$4f_{7/2}-5d_{5/2}$&  4.618&  3.904&  3.671\\
$4f_{5/2}-5g_{7/2}$&  3.250&  2.680&  2.367\\
$4f_{7/2}-5g_{7/2}$&  0.121&  0.100&  0.089\\
$4f_{7/2}-5g_{9/2}$&  4.254&  3.514&  3.117
\end{tabular}
\end{ruledtabular}
\end{table}

We solve the core RPA equations iteratively. 
In our calculations, we set the number of core iteration to 10
to save computation time; for 
convergence to machine accuracy, about 50 iterations  are needed
at low $Z$.
For example, for the $5p_{3/2}-5s$ transition in neutral Ag, first-order
length and velocity matrix elements are 4.30225 and 4.26308,
respectively. The values of the electric-dipole matrix elements
 are given in atomic units, 
$ea_0$. The atomic unit for the corresponding line strength
is $e^2a_0^2$.  
 The corresponding RPA values are
3.77755 and 3.96707 after one iteration; they
become 3.82599 and 3.82636 after 10 iterations.
The final {\em third-order} gauge-independent
results are 3.41726 and 3.41745 for this matrix element in length and velocity
forms, respectively.
\begin{figure*}[tbp]
\centerline{\includegraphics[scale=0.35]{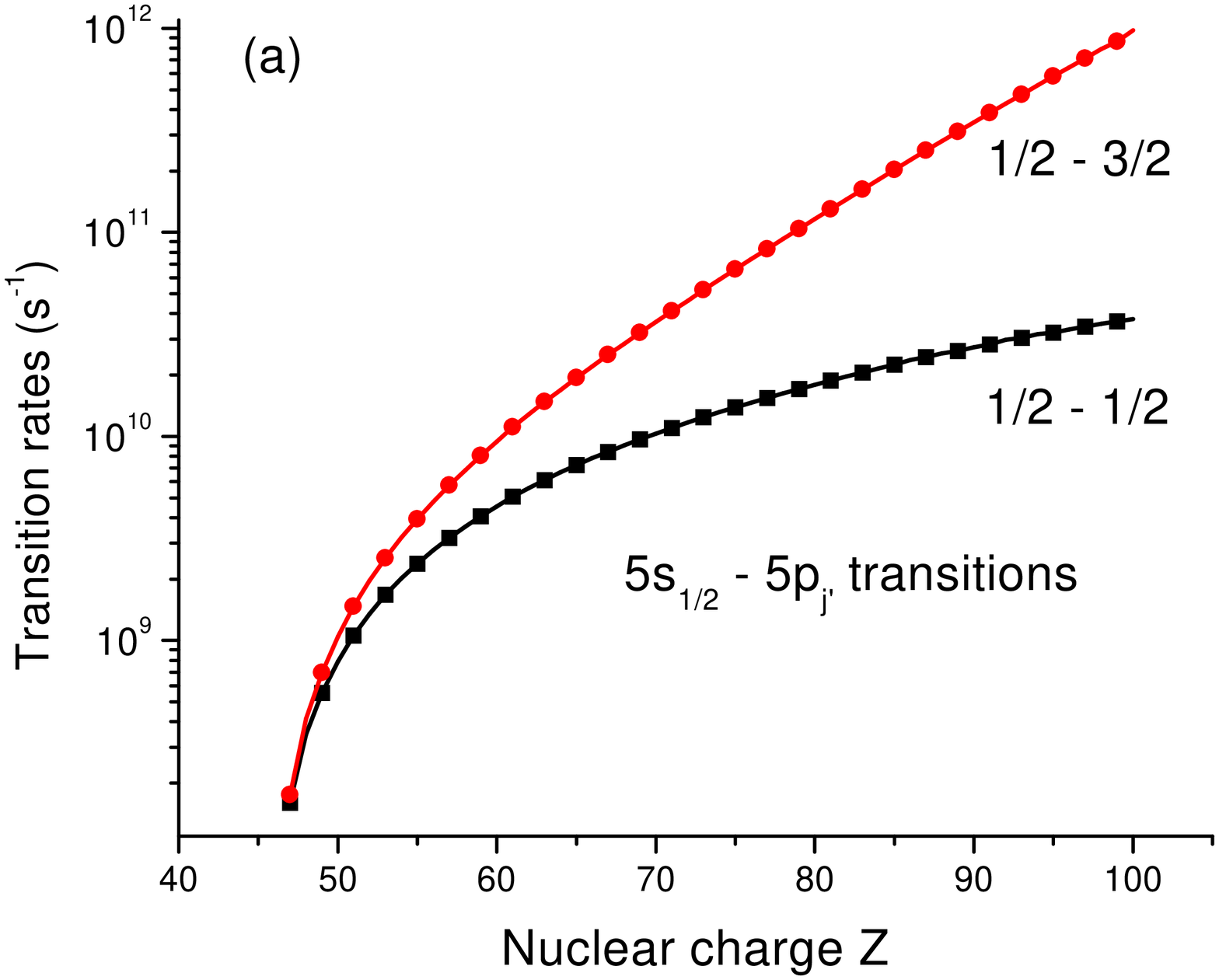}
            \includegraphics[scale=0.35]{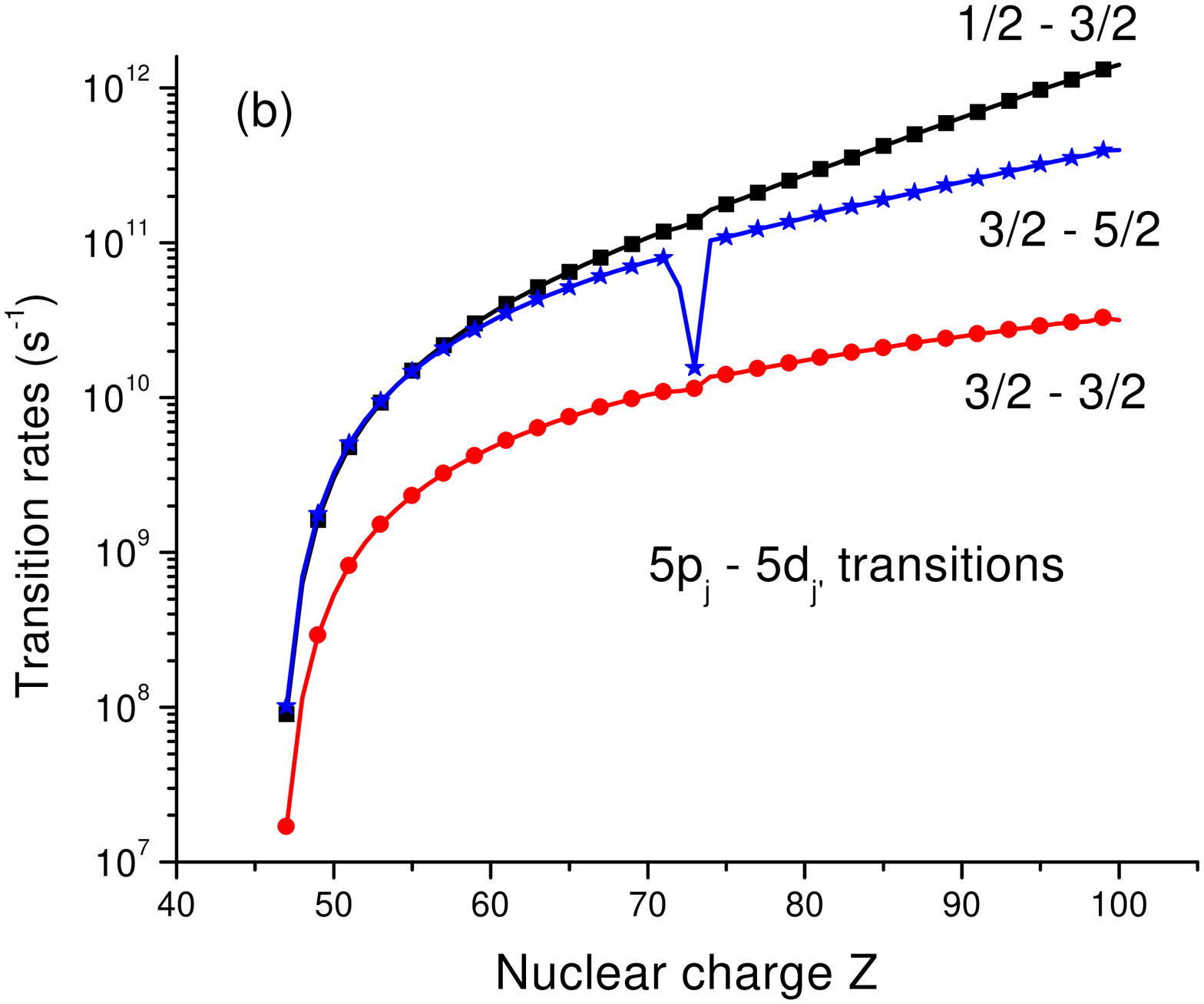}}
\centerline{\includegraphics[scale=0.35]{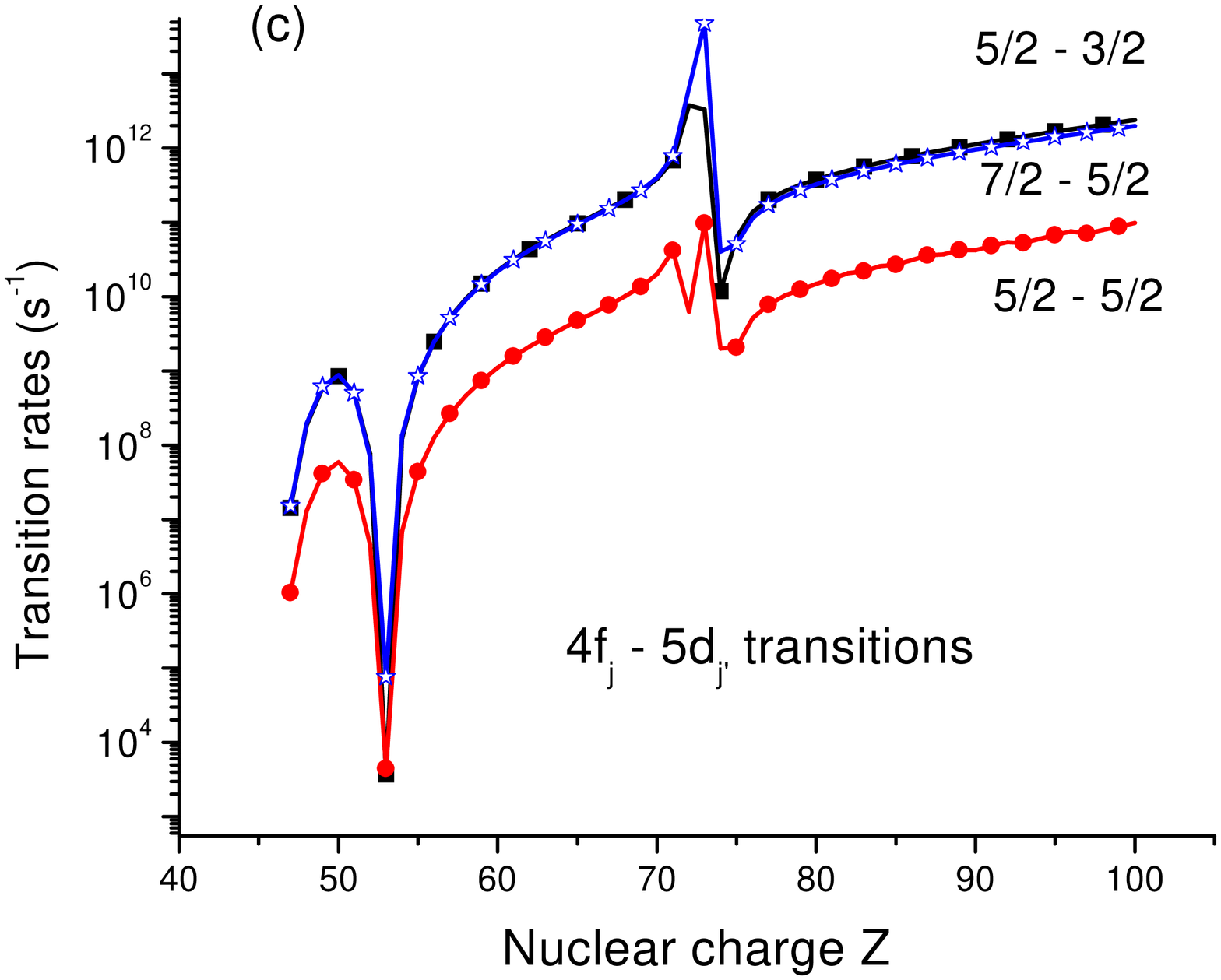}
            \includegraphics[scale=0.35]{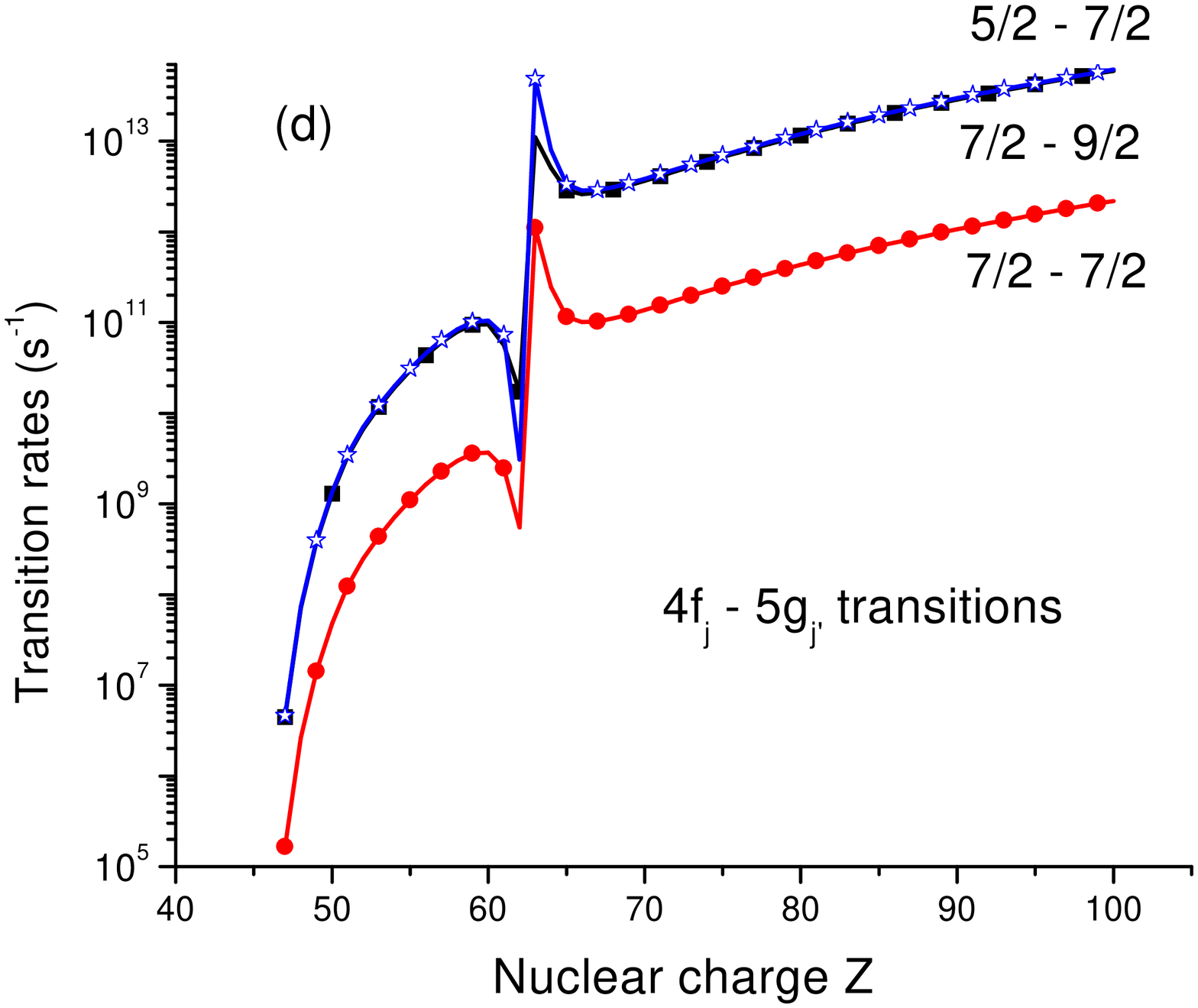}}
\caption{ Transition rates (s$^{-1}$)  in Ag-like ions.}
\label{fig2}
\end{figure*}
\begin{table}
\caption{\label{tab-life} Lifetimes ${\tau}$ in ns
 of the $5l$ and $4f$ levels
 in Cd$^{1+}$, In$^{2+}$,
Sn$^{3+}$, Sb$^{4+}$, Te$^{5+}$,  I$^{6+}$, and Xe$^{7+}$.
The lifetime of the upper level is shown. The corresponding
wavelengths $\lambda$ in \AA~  are also given. In the cases where more
then one transition is allowed the wavelength of the dominant transition
is given. The data are compared with experimental results. }
\begin{ruledtabular}
\begin{tabular}{llllllll}
\multicolumn{1}{c}{Lower}                   &
\multicolumn{1}{c}{Upper}                   &
\multicolumn{1}{c}{$\tau^{(1)}$}            &
\multicolumn{1}{c}{$\tau^{(2)}$}            &
\multicolumn{1}{c}{$\tau^{(3)}$}            &
\multicolumn{1}{c}{$\tau^{\text {expt}}$}   &
\multicolumn{1}{c}{$\lambda^{(3)}$}         &
\multicolumn{1}{c}{$\lambda^{\text {expt}}$}\\
\hline
\multicolumn{8}{c}{Ag~I, $Z$=47}\\
$5s_{1/2}$&$5p_{3/2}$& 7.50&  5.71& 6.97 &6.72$\pm$0.03\footnotemark[1] &3455& 3282\footnotemark[1]\\
$5s_{1/2}$&$5p_{1/2}$& 7.98&  6.24& 7.62 &7.41$\pm$0.04\footnotemark[1] &3562& 3384\footnotemark[1] \\
\multicolumn{8}{c}{Ag-like Cd, $Z$=48}\\
$5d_{5/2}$&$4f_{7/2}$& 5.82& 5.12 & 5.57 &6.7 $\pm$0.2\footnotemark[2] &5417& 5380\footnotemark[2]  \\
$5d_{3/2}$&$4f_{5/2}$& 6.16& 5.41 & 5.90 &6.2 $\pm$0.1\footnotemark[2] &5372& 5338\footnotemark[2]   \\
$5s_{1/2}$&$5p_{3/2}$& 2.32& 2.42 & 2.60 &2.77$\pm$0.07\footnotemark[3]&2170&2145\footnotemark[3]\\
$5s_{1/2}$&$5p_{1/2}$& 2.68& 2.88 & 3.09 &3.11$\pm$0.04\footnotemark[3]&2291&2266\footnotemark[3]\\
$5p_{3/2}$&$5d_{5/2}$& 1.75& 1.44 & 1.67 &1.85$\pm$0.15\footnotemark[3]&2364&2314\footnotemark[3]\\
$5p_{1/2}$&$5d_{3/2}$& 1.95& 1.60 & 1.86 &1.79$\pm$0.11\footnotemark[3]&2243&2195\footnotemark[3]\\
\multicolumn{8}{c}{Ag-like In, $Z$=49}\\
$4f_{7/2}$&$5g_{9/2}$&2.79 &2.52 &2.71 &2.84$\pm$0.30\footnotemark[4]& 4121&4072\footnotemark[4]\\
$5d_{5/2}$&$4f_{7/2}$&1.71 &1.62 &1.74 &1.72$\pm$0.07\footnotemark[4]& 3007&3009\footnotemark[4]\\
$5d_{3/2}$&$4f_{5/2}$&1.78 &1.69 &1.82 &1.70$\pm$0.07\footnotemark[4]& 2969&2983\footnotemark[4]\\
$5s_{1/2}$&$5p_{3/2}$&1.20 &1.42 &1.45 &1.50$\pm$0.15\footnotemark[4]& 1630&1625\footnotemark[4]\\
$5s_{1/2}$&$5p_{1/2}$&1.48 &1.81 &1.84 &1.64$\pm$0.06\footnotemark[4]& 1760&1749\footnotemark[4]\\
$5p_{3/2}$&$5d_{5/2}$&0.58 &0.56 &0.61 &0.58$\pm$0.05\footnotemark[4]& 1507&1488\footnotemark[4]\\
$5p_{1/2}$&$5d_{3/2}$&0.64 &0.61 &0.67 &0.75$\pm$0.06\footnotemark[4]& 1423&1403\footnotemark[4]\\
\multicolumn{8}{c}{Ag-like Sn, $Z$=50}\\
$5d_{5/2}$&$4f_{7/2}$&1.20 &1.27 &1.38  &1.30$\pm$0.20\footnotemark[5]&2266 & 2230\footnotemark[5]\\
$5d_{3/2}$&$4f_{5/2}$&0.98 &1.04 &1.13  &1.25$\pm$0.20\footnotemark[5]&2224 & 2222\footnotemark[5]\\
$5s_{1/2}$&$5p_{3/2}$&0.75 &0.95 &0.95  &0.81$\pm$0.15\footnotemark[5]&1320 & 1315\footnotemark[5]\\
$5s_{1/2}$&$5p_{1/2}$&0.97 &1.27 &1.26  &1.29$\pm$0.20\footnotemark[5]&1444 & 1438\footnotemark[5]\\
$5p_{3/2}$&$5d_{5/2}$&0.29 &0.31 &0.32  &0.45$\pm$0.05\footnotemark[5]&1117 & 1119\footnotemark[5]\\
$5p_{1/2}$&$5d_{3/2}$&0.31 &0.33 &0.34  &0.34$\pm$0.04\footnotemark[5]&1050 & 1044\footnotemark[5]\\
\multicolumn{8}{c}{Ag-like Sb, $Z$=51}\\
$5d_{5/2}$&$4f_{7/2}$&1.77 &2.23 &2.57  &2.5$\pm$0.4\footnotemark[6]  &2268& 2279\footnotemark[6]\\
$5d_{3/2}$&$4f_{5/2}$&1.38 &1.73 &2.00  &2.4$\pm$0.3\footnotemark[6]  &2202& 2217\footnotemark[6]\\
$5s_{1/2}$&$5p_{3/2}$&0.51 &0.68 &0.67  &0.65$\pm$0.12\footnotemark[6]&1108& 1104\footnotemark[6]\\
$5s_{1/2}$&$5p_{1/2}$&0.70 &0.95 &0.92  &0.77$\pm$0.10\footnotemark[6]&1230& 1226\footnotemark[6]\\
$5p_{3/2}$&$5d_{5/2}$&0.18 &0.20 &0.20  &                 &892.1 &     \\
$5p_{1/2}$&$5d_{3/2}$&0.18 &0.21 &0.21  &0.191$\pm$0.020\footnotemark[5]&834.1 &831\footnotemark[5]\\
\multicolumn{8}{c}{Ag-like Te, $Z$=52}\\
$5s_{1/2}$&$5p_{3/2}$& 0.38& 0.510& 0.493& 0.47$\pm$0.03\footnotemark[7] & 952.9& 951\footnotemark[7]\\
$5s_{1/2}$&$5p_{1/2}$& 0.58& 0.738& 0.713& 0.65$\pm$0.04\footnotemark[7] & 1073 &1071\footnotemark[7]\\
$5p_{3/2}$&$5d_{5/2}$& 0.12& 0.140& 0.141& 0.13$\pm$0.03\footnotemark[7] & 745.3& 743\footnotemark[7]\\
$5p_{1/2}$&$5d_{3/2}$& 0.12& 0.146& 0.146& 0.14$\pm$0.04\footnotemark[7] & 693.0& 691\footnotemark[7]\\
\multicolumn{8}{c}{Ag-like I, $Z$=53}\\
$5s_{1/2}$&$5p_{3/2}$& 0.29&  0.39 & 0.38 & 0.35$\pm$0.02\footnotemark[9] &834.7&\\
$5s_{1/2}$&$5p_{1/2}$& 0.43&  0.60 & 0.57 & 0.48$\pm$0.03\footnotemark[9] &954.0&\\
$5p_{3/2}$&$5d_{5/2}$& 0.087& 0.106& 0.105& 0.107$\pm$0.016\footnotemark[8]&641.3&640\footnotemark[8]\\
$5p_{1/2}$&$5d_{3/2}$& 0.090& 0.108& 0.107& 0.120$\pm$0.020\footnotemark[8]&592.9&592\footnotemark[8]\\
\multicolumn{8}{c}{Ag-like Xe, $Z$=54}\\
$5s_{1/2}$&$5p_{3/2}$& 0.23&  0.31&  0.30& 0.33$\pm$0.03\footnotemark[9] &741.0&740.4\footnotemark[1]\\
$5s_{1/2}$&$5p_{1/2}$& 0.35&  0.50&  0.47& 0.50$\pm$0.05\footnotemark[9] &858.6&859.2\footnotemark[1]
\end{tabular}
\end{ruledtabular}
\footnotetext[1] {Ref.~\protect\cite{ag1}} \footnotetext[2]
{Ref.~\protect\cite{cd1}} \footnotetext[3] {Ref.~\protect\cite{cd2}}
\footnotetext[4] {Ref.~\cite{49}} \footnotetext[5]
 {Ref.~\protect\cite{50}} \footnotetext[6] {Ref.~\protect\cite{51}}
\footnotetext[7] {Ref.~\cite{52}} \footnotetext[8]
{Ref.~\protect\cite{53}} \footnotetext[9] {Ref.~\protect\cite{kim}}
\end{table}

The results of our third-order calculations 
are summarized in Table~\ref{tab-osc}, where we list
oscillator strengths for $5s-5p$, $5p-5d$, $4f-5d$, and $4f-5g$
transitions in neutral Ag and low-$Z$ Ag-like ions with $Z = 48 - 60$.

In Table~\ref{tab-s}, we present line strengths for $5s-5p$,
$5p-5d$, $4f-5d$, and $4f-5g$ transitions in Xe$^{7+}$. The
values  calculated
in length form in  first, second, and third
approximations are listed in columns $S^{(1)}$, $S^{(2)}$, and
$S^{(3)}$, respectively. The
difference between second-order values $S^{(2)}$ and third-order
values $S^{(3)}$ is much
smaller than the difference between $S^{(1)}$ and $S^{(2)}$. The
second-order corrections change $S^{(1)}$ by 20 - 50~\%.
The addition of the third-order corrections modifies line strengths
by 5 - 10~\%. The first approximation is just the frozen-core DF 
approximation and the first-order line strengths $S^{(1)}$ 
in Table~\ref{tab-s}
are very close to the earlier DF calculations by \citet{kim}.

\subsection*{$Z$ dependence of transition rates}
Trends of the $Z$ dependence of transition rates are
shown in  Fig.~\ref{fig2}. The $5s-5p$,
$5p-5d$, $4f-5d$, and $4f-5g$
transition rates are shown in Fig.~\ref{fig2}~a, b, c, d,
respectively. All graphs are plotted using second-order data
for consistency. The $Z$ dependences of the transition rates for
$5s-5p$ transitions shown in  Fig.~\ref{fig2}a and two $5p-5d$ transitions
shown in  Fig.~\ref{fig2}b are smooth; however, all
other $Z$ dependences shown in Fig.~\ref{fig2} contain sharp features.
The sharp feature in the curve describing the $5p_{3/2}-5d_{5/2}$ transition
rates (Fig.~\ref{fig2}b) is explained by irregularity in the
curve describing the $5d_{5/2}$ energy shown in Fig.~\ref{fig1}b.
This irregularity in the energy $Z$ dependence was already
discussed in the previous section.

The sharp minima in the region $Z = 52 - 54$ in the curves
describing the $4f-5d$
transition rates shown in Fig.~\ref{fig2}c  are due to
inversion of the order of $4f$ and $5d$ energy levels.
In the region $Z = 52 - 54$ the $4f-5d$
transition energies become very small resulting in the small
transition rates.  The second sharp feature in the curves describing the
$4f-5d$ transition rates shown in Fig.~\ref{fig2}c occurs in the
region $Z$ = 72 - 73 and results from the irregularity in the second-order
correction to the  $4f-5d$ dipole matrix elements.
Below, we describe some details
of the calculation to clarify this issue.

A typical contribution from one of  the second-order RPA
corrections to dipole matrix element ($v-v'$) has the form
\cite{be-tr}
\begin{equation}\label{b5}
D^{{(\rm RPA)}}[v-v']\propto \sum_{nb} \sum_{k}
\frac{D_{nb}X_{k}(vnv'b)}{\epsilon _{n}+\epsilon _{v}-\epsilon
_{v'}-\epsilon _{b}}\, .
\end{equation}
Here, the index $b$ designates a core state and index $n$ designates
an excited state. The numerator is a product of a dipole matrix element 
$D_{nb}$ and a Coulomb matrix element $X_k(vnv'b)$. 
For the special case of the $4f_{5/2}-5d_{3/2}$
transition, the energy denominator for the term in the sum with $n=4f_{5/3}$ and $b=4d_{5/2}$ is 
\begin{eqnarray}
\lefteqn{\epsilon_{n}+\epsilon_{v}-\epsilon_{v'}-\epsilon_{b} }\nonumber \\
&& = \epsilon_{4f_{5/2}}+\epsilon_{4f_{5/2}}-\epsilon_{5d_{3/2}}-\epsilon_{4d_{5/2}} \nonumber \\
&& = -28.0685-28.0685+19.2215+36.8850\nonumber\\
&&=-0.0305.  \label{b8}
\end{eqnarray}
Again, as in the case of the second-order $4d_{3/2}$ energy, there is
a nearly zero denominator when the lowest-order energies 
of the $5d_{3/2}$ and $(4d_{5/2})^{-1}(4f_{5/2})^2$ states are close.
The cause of this irregularity is once again traced to the near degeneracy 
of a single-particle state and a two-particle one-hole state.
The remaining irregularities for $Z>60$ in the curves presented in Fig.~\ref{fig2} 
have similar origins.

\subsection*{Results and comparison for lifetimes}
We calculate  lifetimes of $5l_j$ and $4f_j$ levels in neutral Ag and  in
Ag-like ions with $Z = 48 - 60$ using third-order MBPT results
for dipole matrix elements and
energies. In  Table~\ref{tab-life}, we compare our lifetime data with available
experimental measurements. This set of data includes results for a
limited number of levels in low-$Z$ ions (up to $Z = 54$). We give a more complete comparison
of the transition rates and wavelengths for the
eleven transitions between $5l$ and $4f$ states in Ag-like
ions with $Z = 47 - 60$ including the third-order
contribution in Table III of the accompanying EPAPS document \cite{EPAPS}.
 In Table~\ref{tab-life}, we present our lifetime data $\tau$ calculated
in the lowest-, second-, and third-order approximations.
These results are listed in columns labeled $\tau^{(1)}$, 
$\tau^{(2)}$, and $\tau^{(3)}$, respectively.
The largest
difference between the calculations in different approximations
occurs for $5d_{3/2}$ and $5d_{5/2}$ levels
for $Z$ = 51 and 52 when $5d - 4f$ transition energies
become very small and contributions from the second and third orders
become very important. It should be noted that for some
levels  of neutral Ag and Ag-like  ions with $Z$ = 48 and 49,
$\tau^{(3)}$ agrees better with $\tau^{(1)}$
than with $\tau^{(2)}$. The accuracy of lifetime
measurements is not very high for Ag-like ions, and in some cases
the lowest-order results $\tau^{(1)}$, which are equivalent to the
Dirac-Fock results of \citet{kim} were enough to
predict the lifetimes. The more sophisticated theoretical
studies published recently in Refs.~\cite{chou,martin} were
restricted to  $5s - 5p$ transitions and did not include
wavelength data. In two last columns of Table~\ref{tab-life}, we
compare our theoretical wavelengths, $\lambda^{(3)}$
with experimental measurements, $\lambda^{\text{expt}}$.
In the cases where more
than one transition is allowed, the wavelength of the dominant transition
is given. We find good agreement, 0.01 - 1\%, of our wavelength results
with available experimental data for ions with $Z > 49$.

\section{Conclusion}
In summary,  a systematic  RMBPT study of the energies
of  $5s_{1/2}$, $5p_{1/2}$, $5p_{3/2}$, $5d_{3/2}$,
$5d_{5/2}$, $4f_{5/2}$, $4f_{7/2}$, $5f_{5/2}$, $5f_{7/2}$,
$5g_{7/2}$, and $5g_{9/2}$ states in Ag-like ions is presented.
These energy calculations are found to be in good agreement with existing
experimental energy data and provide a  theoretical
reference database for the line identification.
A systematic relativistic
RMBPT study of reduced matrix elements, line strengths, oscillator
strengths, and transition rates for the 17 possible
$5s-5p$, $5p-5d$, $4f-5d$, and $4f-5g$ electric-dipole
transitions in Ag-like ions throughout the isoelectronic sequence
up to $Z=100$ is conducted.
Both length and velocity forms of
matrix elements are evaluated. Small differences between length and
velocity-form calculations, caused by the nonlocality of the DF potential,
are found in second order. However, including third-order
corrections with full RPA leads to complete
agreement between the length- and velocity-form results.

We believe that our energies and transition rates will be useful in analyzing existing
experimental data and planning new experiments. There remains a
paucity of experimental data for many of the higher ionized
members of this sequence, both for term energies and for
transition probabilities and lifetimes.
\begin{acknowledgments}
The work of W. R. J. and I. M. S. was supported in part by National
Science Foundation Grant No.\ PHY-01-39928. U.I.S. acknowledges
partial support by Grant No.\ B516165 from Lawrence Livermore
National Laboratory.
\end{acknowledgments}

\end{document}